\documentclass[runningheads]{svmult}

\usepackage{makeidx}   % allows index generation
\usepackage{graphicx}  % standard LaTeX graphics tool
\usepackage{subeqnar}  % subnumbers individual equations
\usepackage{multicol}  % used for the two-column index
\usepackage{physprbb}  % modified textarea for proceedings,
\makeindex             % used for the subject index
\begin{document}
\title*{ELT Observations of Supernovae at the Edge of the
Universe}
\toctitle{Focusing of a Parallel Beam to Form a Point
\protect\newline in the Particle Deflection Plane}
% allows explicit linebreak for the table of content
%
%
\titlerunning{Simulated Observations of SNe}
% allows abbreviation of title, if the full title is too long
% to fit in the running head
%
\author{Massimo Della Valle\inst{1}
\and Roberto Gilmozzi\inst{2}
\and Nino Panagia\inst{3}
\and Jacqueline Bergeron\inst{4}
\and Piero Madau\inst{5}
\and Jason Spyromilio\inst{2}
\and Philippe Dierickx\inst{2}}
\authorrunning{Della Valle et al.}
% if there are more than two authors,
% please abbreviate author list for running head
%
%
\institute{INAF-Arcetri Astrophysical Observatory, Largo E. Fermi 5, 50125, Firenze, Italy 
\and European Southern Observatory, 3107 Alonso de Cordova, Santiago, Chile
\and ESA/Space Telescope Science Institute, 3700 San Martin Drive, Baltimore, MD 21218, USA 
\and Institut d'Astrophysique de Paris - CNRS, 98bis Boulevard Arago, 75014 Paris, France 
\and Department of Astronomy and Astrophysics, University of California, 1156 High Street, 
Santa Cruz, CA 95064, USA 
\and European Southern Observatory, Karl-Schwarzschild-Strasse 2, Garching D-85748, Germany }

\maketitle              % typesets the title of the contribution

\begin{abstract}

We discuss the possibility of using Supernovae as tracers of the star
formation history of the Universe for the range of stellar masses
$\sim 3-30$ M$_\odot$ and possibly beyond. We simulate the
observations of 350 SNe, up to $z\sim 15$, made with OWL (100m)
telescope.

\end{abstract}

\section{Introduction}

The detection and the study of Supernovae (=SNe) is important for at
least two reasons: {\sl 1.} the use of {\sl local} SNe (both type Ia
and II) as `calibrated' standard candles (Phillips 1993, Hamuy et
al. 2001, Hamuy \& Pinto 2002) provides a direct measurement of the
expansion rate of the Universe H$_\circ$, and their detection at
$z>0.3$ allows to measure its deceleration parameter q$_\circ$ and to
probe different cosmological models (Perlmutter et al. 1998, 1999;
Riess et al. 1998); {\sl 2.} the evolution of the cosmic SN rate
provides a direct measurement of the cosmic star formation rate
(SFR). Indeed the rate of core-collapse SN explosions (SN II, Ib/c) is
a direct measurement of the death of stars with masses in the range
8-30 M$_\odot$ (although it is still debated if stars more massive
than 30 M$_\odot$ could make "normal" type II/Ibc SNe, or rather
collapse forming a BH with no explosion at all, or even make a
different kind of explosion like GRBs; see, e.g., Heger et al. 2001).
Similarly, type Ia SNe may provide the history of star formation of
moderate mass stars, 3-8 $M_\odot$, i.e. their most likely progenitors
provided that the SN Ia explosion process is unambiguously identified
(e.g. Mannucci, Della Valle \& Panagia 2005). In addition the
evolution of the SN-Ia rate with redshift helps to clarify the nature
of their progenitors (e.g. Madau, Della Valle \& Panagia 1998, Dahlen
et. al 2004, Strolger et al. 2004).

\section{SNe as tracers of star formation}

For a Salpeter Initial Mass Function (IMF) with an upper cutoff at 100
M$_\odot$ we find that half of all SNII is produced by stars with
masses between 8 and 13 M$_\odot$ and half of the mass in SN-producing
stars is in the interval of mass $\sim 8-22$ M$_\odot$. A main
sequence star with 13 M$_\odot$ has approximately a luminosity of 8000
L$_\odot$ and a temperature of 22000 K and one with 22 M$_\odot$ has
L$\sim 3.5\times 10^4 L_\odot$ and T$_{eff} \sim 27000$ K. This means
that more than half of the SN producing stars are rather poor sources
of ionizing photons and of UV continuum photons, while the bulk of the
UV radiation, both in the Balmer and in the Lyman continuum is
produced by much more massive stars. Actually, starburst models
(e.g. Leitherer et al. 1999) suggest us that stars above 30 M$_\odot$
produce 90\% of the Lyman continuum photons and 70\% of the
912-2000\AA~ UV continuum.  It follows that the bulk of the UV
radiation both in the Balmer and Lyman continuum is produced by stars
more massive than 30$M_\odot$, so that the H$\alpha$ and UV fluxes
measure {\bf only} the very upper part of the IMF (stars with masses
$>30_\odot$ represent only 13\% of the total number of stars with
masses larger than $>8M_\odot$). Therefore, H$\alpha$ and UV flux are
not ideal indicators of star formation rate because: {\sl a)} they
require a huge extrapolation to lower masses and {\sl b)} the
extrapolation depends on the value of minimum mass to make a SNII,
M$_{up}$, which is not well known and may be different in different
environments (Bressan et al. 2002) or at different redshifts (Heger et
al.  2001). On the other hand, SNe can provide a measurement of the
star formation rate which is: {\sl i)} independent of other SFR
determinations; {\sl ii)} more direct because the IMF extrapolation is
appreciably smaller; {\sl iii)} more reliable because it is based on
counting SN explosions rather than relying on identifying and
measuring the sources of ionization or of UV continuum.  A possible
drawback for this approach is the fact that a significant fraction of
SNe may be missed because of extinction in their parent galaxies
(Mannucci et al. 2003).

\section{The Ingredients of the Simulation}

Our simulation is based on a number of assumptions:

{\sl 1. ELT performance}. We have assumed ELT= OWL, i.e. a 100m
telescope. The imaging and spectroscopic limits are a function of the
S/N ratio as derived from the ELT exposure time calculator
(www-astro.physics.ox.ac.uk/~imh/ELT).

{\sl 2. OWL field}. A reasonable extrapolation of the current
technological standards let us presume that diffraction limited
observations of a $2' \times 2'$ field, corrected with adaptive optics
(K band), is within the technological possibilities of the next decade
(Ragazzoni, private communication).

{\sl 3. Number of SNe expected in a single OWL frame}. Based on the
results obtained by Madau et al. (1998), Miralda Escud\'e \& Rees
(1997), Mackey et al. (2003) and Weinmann \& Lilly (2005) we
estimate an observed rate of up to 7 SNe/yr  per OWL field.
Here we are including the very powerful Pop~III SNe that are expected
to be produced by pair-creation in zero-metallicity massive stars, in
the range $140-260M_\odot$ (Heger et al. 2001).

{\sl 4. Spectral energy distribution (=SED) of SNe} (see Fig. 1). We
have assumed as templates for type Ia and II : SN 1992A, SN 1999em and
SN 1998S (Panagia 2003, Riess et al. 2004).  The SED for SNe
originating from Pop~III stellar population has been obtained from
Heger et al. (2001).

\begin{figure}[a]
\begin{center}
\includegraphics[width=0.6\textwidth]{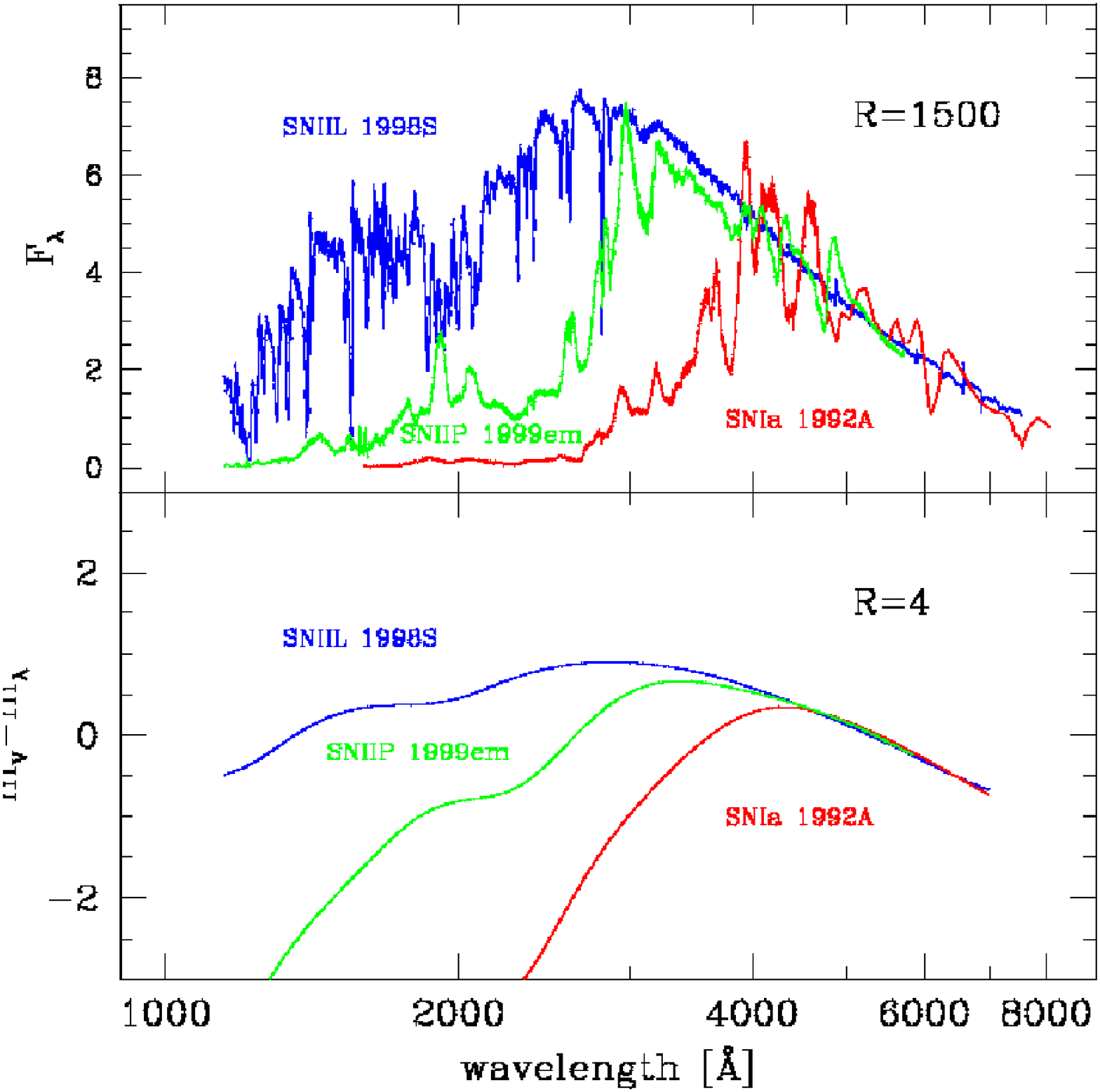}
\end{center}
\caption[]{Spectral Energy Distribution adopted for type I and type II
SNe. [Adapted from Panagia 2003]}
\label{eps1}
\end{figure}

{\sl 5. Observational strategy}. It is based on the control time
methodology (Zwicky 1938). The morphology of the light curves and the
absolute magnitudes at maximum of SNe have been obtained from Barbon,
Ciatti \& Rosino (1973), Dogget \& Branch (1985), Patat et al. (1994)
and Saha et al. (2001).

{\sl 6. Distribution of SNe into spectroscopic types}.  These data
have come from: Mannucci et al. (2004), Della Valle et al. (2005),
Mackey et al. (2003), and Cappellaro et al. (1997). We have estimated
about 63\% of SNe to be of type II (a few SNe-II are expected to be
observed during the UV shock breakout, that at z$\sim 7-8$ should last
about 4-5 days in the observer rest frame), about 20\% type Ia,
16\% type Ib/c, and about 1\% SNe from Pop~III.

7. Cosmology ($\Omega_m=0.3$, $\Omega_\Lambda=0.7$) (e.g. Knop et al.
2003, Riess et al. 2004)

\section{Redshift Thresholds}

Fig. 2-5 represent the spectroscopic templates for SNeI-a, SNeII, and
Pop~III at different redshifts as viewed in the observer
rest-frame. The three solid lines denote the fluxes, at different
resolution factors, corresponding to S/N=10 for OWL exposures of
$10^5$ sec.  The dashed line is the threshold (R=5) for JWST (Panagia
et al. 2003).  Scaling the data in Figures 2-5 reveals that SNe are
well detectable with OWL in 1h exposure. SNe-Ia are detectable up to
$z\sim 5$ while SNe-II (bright) up to $z\sim 7-8$. Pop~III SNe, which
in principle would be easily detectable up to $z\sim 20$ (Heger et
al. 2001) are unlikely observable beyond $z\sim 15$, due to the Lyman
forest absorption.

\begin{figure}[b]
\begin{center}
\includegraphics[width=0.8\textwidth]{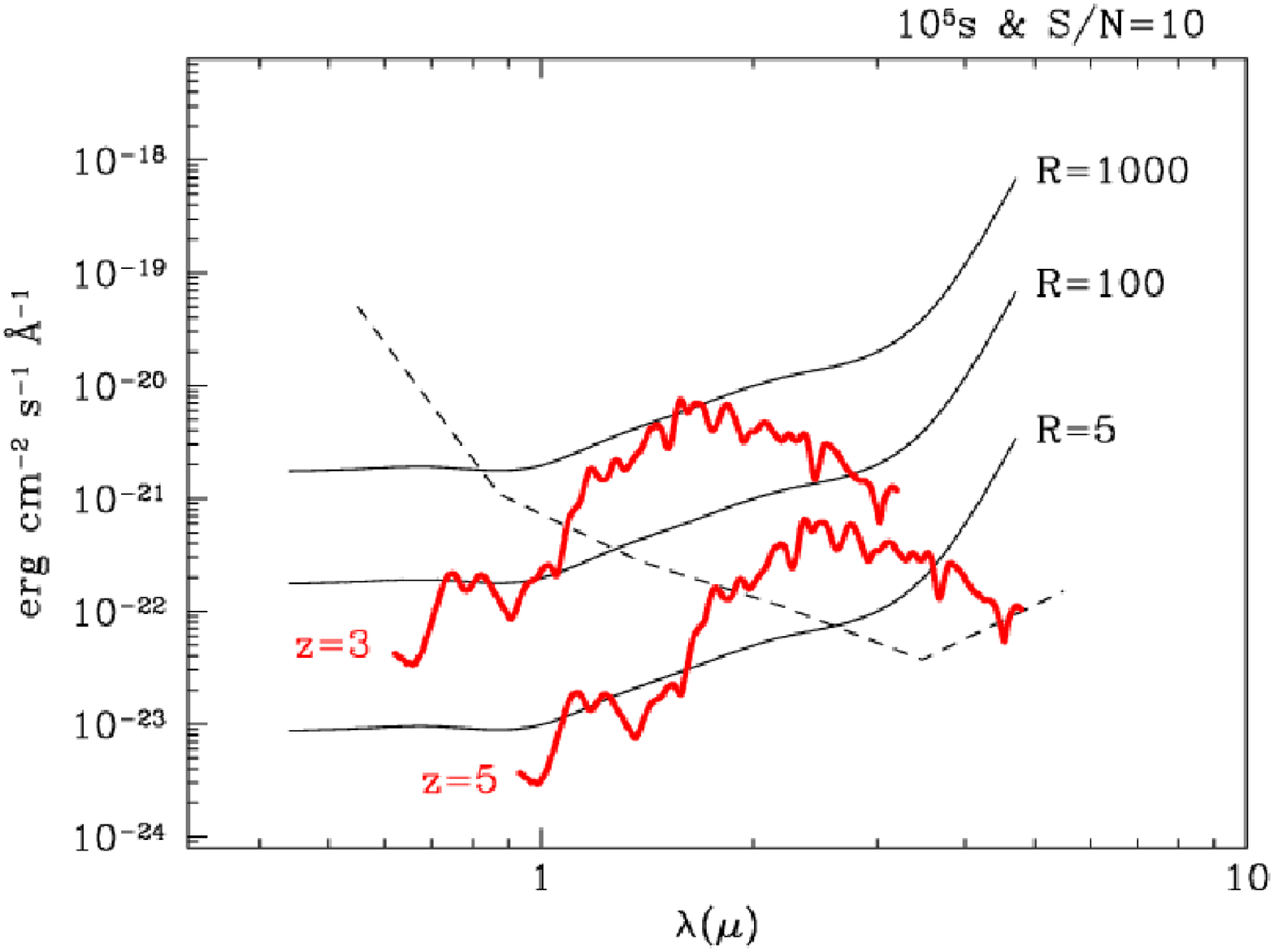}
\end{center}
\caption[]{The spectroscopic template for type Ia SNe as viewed in
the rest-frame of the observer, at different redshifts. The three
solid lines denote the fluxes corresponding to a S/N=10 for OWL
exposures of $10^5$ s at different resolutions. The dashed line is
the threshold (R=5) for JWST.}
\label{eps2}
\end{figure}

\begin{figure}[c]
\begin{center}
\includegraphics[width=0.8\textwidth]{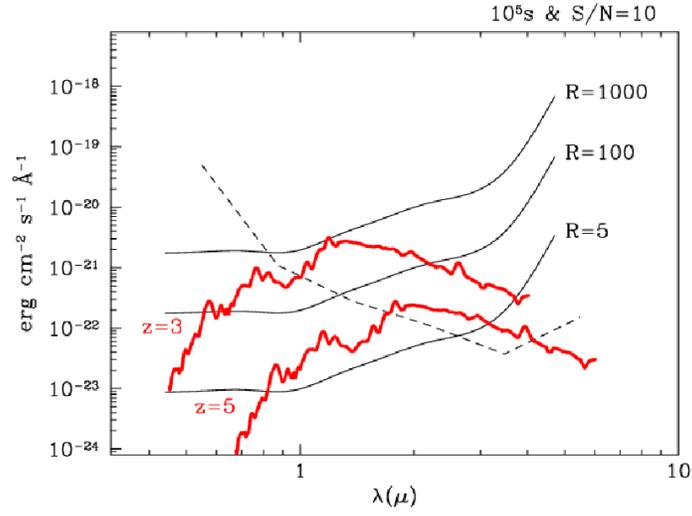}
\end{center}
\caption[]{The same as Fig. 2 for SNe-II (regular).}
\label{eps3}
\end{figure}

\begin{figure}[d]
\begin{center}
\includegraphics[width=0.8\textwidth]{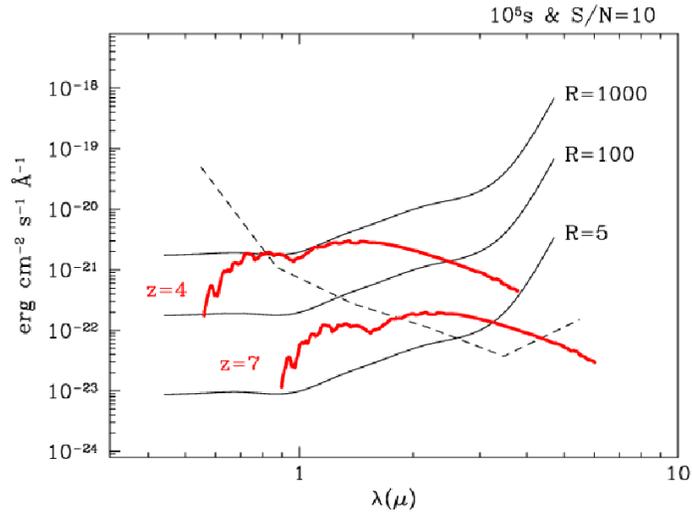}
\end{center}
\caption[]{The same as Fig. 2 for SNe-II (bright).}
\label{eps4}
\end{figure}

\begin{figure}[e]
\begin{center}
\includegraphics[width=0.8\textwidth]{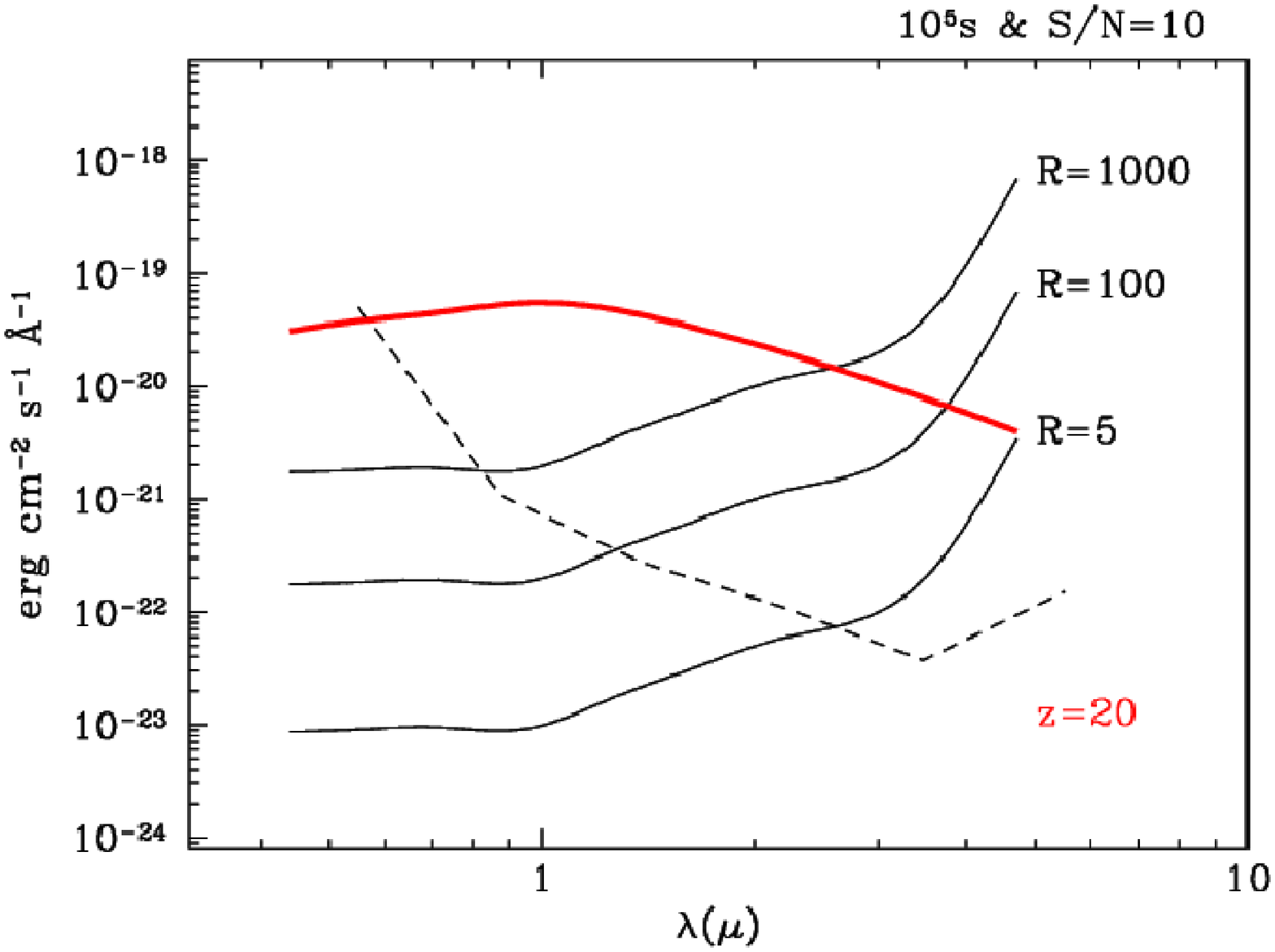}
\end{center}
\caption[]{The same as Fig. 2 for SNe originating from Pop~III
stellar population. The template has been derived from Heger et al. 2001
(their Fig. 3).}
\label{eps5}
\end{figure}

\section{The Simulation} 

An ELT project devoted to the study the evolution of the cosmic SN
rate up to $z\sim 15$ requires an important (although not huge, as we
shall see) investment of telescope time. We plan to carry out the SN
search on 50 OWL fields in the J, H and K bands (1 hour each) at 4
different epochs over an interval of time of 1 year. This strategy is
justified by the following considerations.  The typical light curve
width around maximum light is 15-20 days in the SN rest frame, and
most SNe will occur at $z<5$ (see Fig. 2-5). Therefore, the light
curve widths in the observer rest frame correspond to about 100-120
days, so that 4 exposures obtained 3 months apart will cover all
events occurring within 1 year. In addition we need 3 more epochs (1h
each) in the K band for the photometric follow-up and 1 spectroscopic
epoch (4h for each SN discovered at $z<5$) for the spectroscopic
classification. At very high redshifts, z $\sim 7-8$, only bright type
II SNe may be used as standard candles (due to their strong UV
emission, while SNe Ia are basically blind below $2400\AA$~ and
therefore they are detectable `only' up to z$\sim 5$).  Bright type II
SNe can be standardized via ``expanding photosphere method'' (Hamuy et
al. 2001) or alternatively via expansion velocity vs. bolometric
luminosities relationship at the plateau stage (Hamuy \& Pinto
2002). This task can be accomplished by securing a second and possibly
a third spectroscopic epoch, and this requires about 200 hours of
observing time. Finally, we need 4h for each Pop~III SN to obtain
medium/high resolution spectroscopy.

\begin{figure}[f]
\begin{center}
\includegraphics[width=0.8\textwidth]{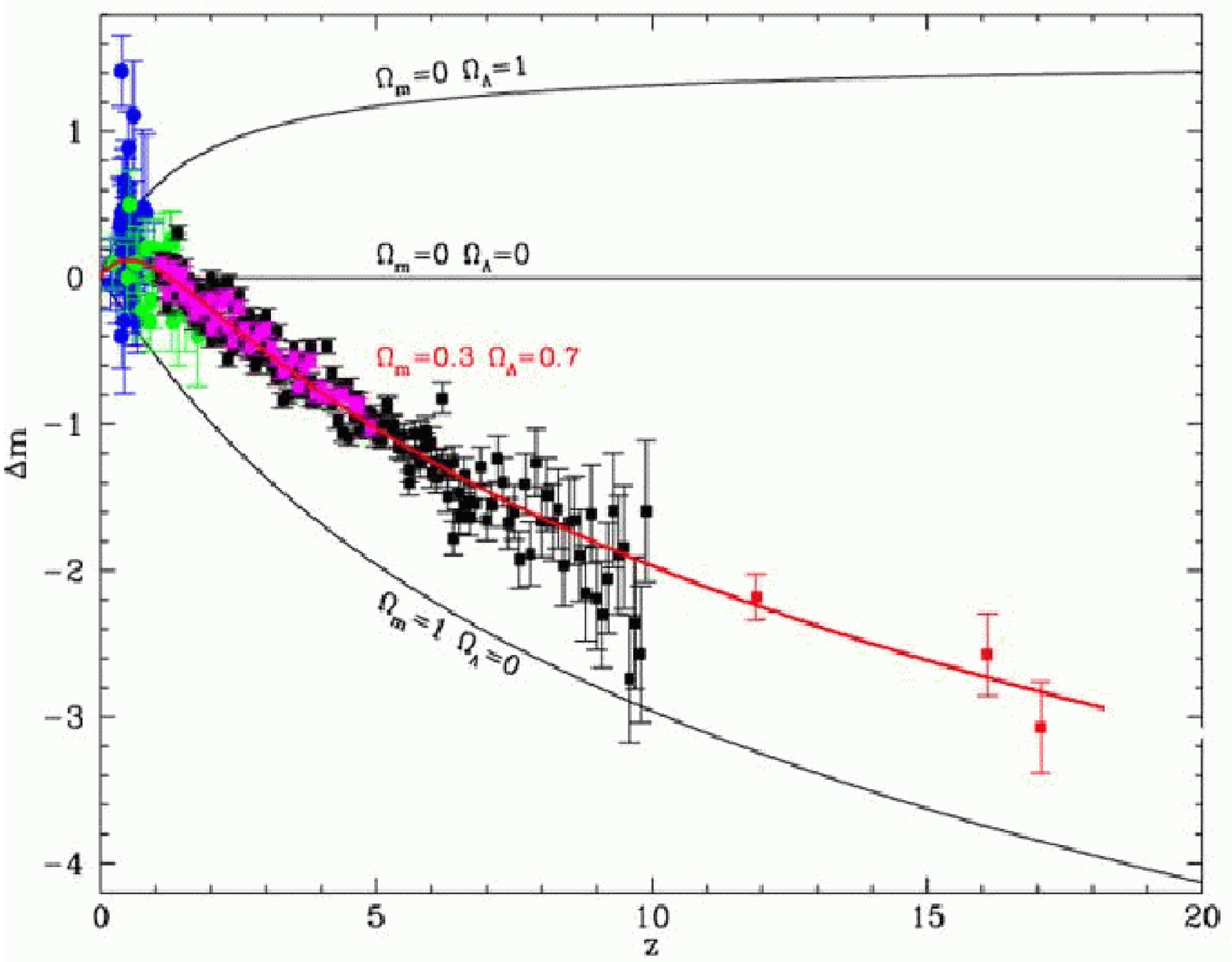}
\end{center}
\caption[]{Hubble diagram for the simulated ELT observations of SNe.
Pink, black and red dots represent type Ia, type II (+Ib/c) SNe and
SNe from Pop~III stellar population. Blue and green dots are `real'
SNe observed with ground based telescopes and HST, respectively.}
\label{eps6}
\end{figure}

In summary, this programme requires a ``grand total'' of 1270h, of
which 600h are for the SN search, 150h for the photometric follow-up
(K band) and 200h for the spectroscopy of type Ia, Ib/c and II SNe,
200 additional hours for the second and third spectroscopic epochs of
bright type II SNe and finally 120h for the spectroscopy of Pop~III
SNe. All of this corresponds to about 160 nights, which will allow us
to study about {\bf 350 } SNe. This is certainly an important
investment in terms of telescope time, however we note that this
corresponds to about three times as much the size of a current HST
Treasury programme (450 orbits) and it is comparable with the
requirements of the SNAP (now Joint Dark Energy Mission) project,
which is expected to study about 2000 SNe Ia (at $z<2$) in 2 years
(Aldering et al. 2002). In Fig. 6 we show the `virtual' SNe discovered
by OWL: pink dots are type Ia SNe, black dots type II (+Ib/c), blue
and green dots are Ia SNe `actually' discovered by ground based
telescopes (Perlmutter 1998, 1999; Riess 1998; Knop et al.  2003,
Tonry et al. 2003) and from HST (Riess et al. 2004).  The SNe have
been distributed around the track $\Omega_m=0.3$, $\Omega_\Lambda=0.7$
after taking into account the intrinsic dispersion of the peak of the
luminosity of type Ia and II SN populations, while the photometric
errors have been derived from the S/N ratio that has been computed for
each simulated observation. Red dots represent SNe from Pop~III
star population. The explosion rate for Pop~III SNe was taken from
Mackey et al. (2003).  However, recently Weinmann \& Lilly (2005)
found that this rate may be too high and it should be decreased by an
order of magnitude. In this paper we assume as an estimate of the rate
of Pop~III SNe (up to z$\sim 15$) in the $2\times 2$ arcmin$^2$ OWL
field: R$_{SNe-III}\simeq f \times d \times 5 \times 10^{-2}$/yr SNe,
being $f$ an efficiency factor ranging between 1 (Mackey et al. 2003)
and 0.1 (Weinmann \& Lilly 2005) and $d$ the duration (in years) of
the SN peak. According to Heger et al. (2001, their Fig. 3) $d$ is
about 1 (at 2-3$\mu$, in the observer rest-frame) therefore, one
should be capable to observe during the OWL survey about $f\times 5 \times
10^{-2} \times 50$ fields $\simeq f\times 3$ SNe at any time.

\section{Conclusions}

In this paper we have argued that SNe can be used as profitable
tracers of cosmic star formation for a number of reasons: {\it i)} the
determinations of the SFR based on SN measurements are independent of
other possible determinations, {\it ii)} SNe provide a more direct
diagnostics than the UV luminosity or the H$\alpha$ line emission
because the IMF extrapolation is much smaller and {\it iii)} SNe are a
more reliable source of information because it is based on a simple
count of individual SN explosions rather than relying on identifying
and measuring the source of ionization (if using H-alpha flux) or the
source of UV continuum.  In addition, by studying SNe at high
redshifts {\it iv)} we can learn to what extent the IMF was more
skewed toward massive stars (relatively to a normal Salpeter's) in low
metallicity environments, and {\it v)} we can study the properties of
the progenitors of the primordial Gamma Ray Bursts (=GRBs) (given the
growing evidence for the existence of an association between core
collapse SNe with the long duration GRBs, see Della Valle 2005 and
references therein).

The results of our simulation indicate that $\sim 350$ SNe can be
studied up to $z\sim 15$ within a reasonable amount of telescope time
(about 160 nights). This pilot programme has been conceived to study
the history of the cosmic star formation rate, nevertheless a number
of important by-products are at hand: such as {\it vi)} to disentangle
cosmological models alternative to $\Lambda$, {\it vii)} to clarify
the nature of the progenitors of type Ia SNe {\it viii)} to probe the
physical properties of the ISM and IGM at $z>10$ through high spectral
resolution (R$\sim 10^4$) of Pop~III SNe (feasible with 50--100m
telescope) {\it ix)} to explore the metal enrichment of the IGM at
early epochs (up to $z\sim 4$) via observations of bright type II and
Ia SNe at a resolution of $R\sim 1000$ (feasible with 50-100m
telescope).

%INDEX%%%%%%%%%%%%%%%%%%%%%%%%%%%%%%%%%%%%%%%%%%%%%%%%%%%%%%%%%%%%%%%
% Please check with the editor of your book whether he plans to
% include a "mutual" subject index - if so, please code your entries
% in the standard syntax. For your own purposes you may print your
% "personal" index by using the following commands:
%
%\clearpage
%\addcontentsline{toc}{section}{Index}
%\flushbottom
%\printindex
%%%%%%%%%%%%%%%%%%%%%%%%%%%%%%%%%%%%%%%%%%%%%%%%%%%%%%%%%%%%%%%%%%%%%

\end{document}